\documentclass[11pt]{article}
\usepackage{amsmath,amssymb,amsbsy,geometry,graphicx,enumerate,psfrag,subfigure}

\newcommand{\got}[1]{\mathcal{#1}}

\newcommand{\mymin}[1]{\mathop{\textrm{min}}\limits_{#1}}

\newcommand{\pd}{\partial}
\newcommand{\Hn}[2]{H_{#1}^{(#2)}}

\newcommand{\vc}[1]{{\boldsymbol{#1}}}

\newcommand{\fbrl}{\Biggl\{}
\newcommand{\fbrr}{\Biggr\}}

\def\approxF{\hbox{\space \raise-2mm\hbox{$\textstyle      \approx \atop \scriptstyle \mathcal{F} \ll 1$} \space}}

\newcounter{mylistcounter1}

\newcounter{mylistcounter2}

\newcounter{mylistcounter3}
\newenvironment{mylist3}{\begin{list}{Region (\Roman{mylistcounter3})~}
	{\usecounter{mylistcounter3}\setlength\labelwidth{1cm}\setlength\leftmargin{1cm}\setlength\itemsep{5pt}}}
{\end{list}}

\begin{document}

\title{Analytical approximations for low frequency band gaps in periodic arrays of elastic shells}

\author{%
  Anton Krynkin$^1$, Olga Umnova$^2$, Shahram Taherzadeh$^3$ and Keith Attenborough$^3$\\
	{\small $^1$ School of Engineering, Design and Technology, University of Bradford, Bradford, UK }\\
	{\small $^2$ Acoustics Research Centre, The University of Salford, Salford, Greater Manchester, UK}\\
	{\small $^3$ Department of Design Development Environment and Materials, The Open University, Milton Keynes, UK}\\
	{\small email: a.krynkin@bradford.ac.uk, O.Umnova@salford.ac.uk}
}

\date{\today}

\maketitle

\begin{abstract}

This paper presents and compares three analytical methods for calculating low frequency band gap boundaries in doubly periodic arrays of resonating thin elastic shells. It is shown that both lattice sum expansions in the vicinity of its poles and self consistent scheme could be used to predict boundaries of low-frequency band gaps due to axysimmetric ($n=0$) and dipolar ($n=1$) shell resonances. The accuracy of the former method is limited to low filling fraction arrays, however the application of the matched asymptotic expansions could significantly improve approximations of the upper band gap boundary due to axysimmetric resonance. The self-consistent scheme is shown to be very robust and gives reliable results even for dense arrays with filling fraction higher than $40\symbol{37}$. The results are used to predict the dependence of the position and width of the low frequency band gap on the properties of shells and their  periodic arrays.

\end{abstract}

\section{Introduction}

Periodic and random arrangements of scatterers with low frequency resonances in air have attracted a significant amount of attention recently due to their applications as locally resonant acoustic metamaterials \cite{Liu,Guenneau,Fokin}. Different types of scatterers have been shown to possess these types of resonances, among them most notable are split rings i.e. 2D analogues of Helmholtz resonators \cite{MovchanP1}.

It has been demonstrated recently that periodic arrays of thin elastic shells in air can possess multiple resonant band gaps in the frequency range below the first Bragg band gap \cite{SUOU,Kosevich}. Data from experiments with finite arrays of elastic shells \cite{SUOU} have shown that the insertion loss (IL) peak corresponding to the axisymmetric ($n = 0$) resonance is comparable to that for the first Bragg band gap but at a lower frequency. The IL peak corresponding to the $n = 1$ resonance also appears at a lower frequency but is less strong.

The position and width of the complete band gaps can be obtained using standard numerical and semi-analytical techniques \cite{Movchan,Nicorovici}. However it would be attractive to have analytical or semi-analytical expressions allowing approximate estimates of band gap boundaries. This would help to better understand their dependence on scatterer and array parameters and hence facilitate the design of resonant metamaterials with desired properties.

In this paper several methods allowing simple estimations of resonant band gaps boundaries in periodic arrays of elastic shells are presented. In Section~\ref{Foldy} the dispersion relation and expressions for lower and upper band gap bounds are derived using expansion of the lattice sum within the vicinity of its pole~\cite{McIver1}. The derived dispersion relation is similar to that obtained with the Foldy's approximations \cite{Martin}. This method allows analytical approximations of two resonant band gaps ($n=0$ and $n=1$). In Section~\ref{MAE} matched asymptotic expansions are used to improve approximations for the upper boundary of a band gap associated with the axysimmetric ($n=0$) resonance. A self-consistent method, also known as coherent potential approximation (CPA) \cite{Sheng,Berryman}, is used to derive simple expressions for effective density and bulk modulus of the array and hence to obtain a dispersion relation in Section~\ref{CPA}. The band gaps due to $n=0$ and $n=1$ resonances are determined by finding frequency ranges where effective density or bulk modulus (respectively) are negative. The results of all techniques are compared with each other and with numerical model predictions. In Section~\ref{results} the derived approximations are applied to estimate the width of the band gaps with respect to the variation in elastic shell radius, its thickness, lattice cell size (i.e. lattice constant) and Young's modulus of the elastic shell.

\section{Foldy's type formula}\label{Foldy}

First consider an eigenvalue problem stated for acoustic wave propagation through a doubly periodic array of circular scatterers $C_j$, $j=0,1,2...$. The scatterers are arranged in an infinite lattice $\Lambda$ with lattice constant $L$. The acoustic environment is characterised by its density $\rho_o$ and sound speed $c_o$. In this section the origins of Cartesian coordinates $\vc{r}=(x,y)$ and polar coordinates coincide with the center of scatterer $C_0$ in the primary cell. Centre $O_j$ of each scatterer $C_j$ in the $j$th cell of lattice $\Lambda$ is defined by the position vector 
\begin{equation}
\label{ps_directvector}
		R_j = n_1 \vc{a_1} + n_2 \vc{a_2},\quad n_1,\,n_2\in\mathbb{Z}, 
\end{equation}
where $\vc{a_1}$ and $\vc{a_2}$ are the fundamental translation vectors. The solution in acoustic medium $p(\vc{r})$ satisfies the Helmholtz equation,
\begin{equation}
\label{ps_Helmholtz}
	\Delta p + k_o^2 p = 0,
\end{equation}
where $\displaystyle{\Delta = \frac{1}{r}\frac{\pd}{\pd r}\left(r\frac{\pd}{\pd r}\right)+\frac{1}{r^2}\frac{\pd^2}{\pd \theta^2}}$, $k_o=\omega/c_o$ is the wavenumber defined as a ratio between angular frequency $\omega$ and sound speed $c_o$. Throughout this paper the time-harmonic dependence is taken as $exp(-i \omega t)$.

Solution $p(\vc{r})$ is subject to the boundary conditions specified later in this paper and the Floquet-Bloch conditions, also referred to as quasi-periodicity conditions, that are
\begin{equation}
\label{ps_Bloch}
    p(\vc{r}+\vc{R}_j) = e^{i\vc{\beta}\vc{R}_j}\, p(\vc{r});
\end{equation}
where $\vc{r}$ is the position vector of a field point, $\vc{R}_j$ is the position vector of each scatterer centre and $\vc{\beta}=(q_1,q_2)$ is a given wave vector. The components of the wave vector $\vc{\beta}$ can also be defined in polar coordinates as $q_1=\beta\cos\tau$ and $q_2=\beta\sin\tau$.

The general solution of the Helmholtz equation \eqref{ps_Helmholtz} is given by
\begin{align}
\label{ps_ABgensol}
 	p(r,\theta) = \sum_{n=-\infty}^{+\infty} \left[B_{n}^J J_n(k_o r) + B_{n}^Y Y_n(k_o r)\right] e^{i n\theta}.
\end{align}
Applying the quasi-periodicity conditions \eqref{ps_Bloch} together with the boundary conditions imposed on the surface of each scatterer to equation \eqref{ps_ABgensol} results in the dispersion relation referred to as the Rayleigh Identity \cite{Movchan}
\begin{align}
\label{ps_RI}
		B_n^J - \sum_{p=-\infty}^{+\infty}(-1)^{p-n}\sigma^Y_{p-n}(k_o,\vc{\beta})Z_p B_p^J = 0,\, n\in\mathbb{Z}
\end{align}
where factor $Z_n$ is defined by boundary conditions on scatterer's surface. For example, zero normal velocity at the surface of the rigid scatterers of radius $a$ gives 
\begin{align}
\label{ps_rigidBC}
		Z_n=-\frac{J'_n(k_o a)}{Y'_n(k_o a)}
\end{align}
with prime denoting derivative with respect to polar coordinate $r$. The lattice sum $\sigma_n^Y(k_o,\vc{\beta})$ in equation \eqref{ps_RI} can be derived in the lattice $\Lambda$ as
\begin{align}
\label{ps_lsum}
    \sigma_{n}^Y(k_o,\vc{\beta}) = \sum_{\vc{R}_j \in \Lambda\setminus\{\vc{0}\}}  e^{i\vc{\beta}\vc{R}_j} Y_n(k R_j) e^{i n \alpha_j},
\end{align}
where representation $\vc{R}_j = R_j(\cos\alpha_j,\sin\alpha_j)$ has been used. 

The lattice sum can also be derived in a reciprocal lattice $\Lambda^\star$ defined by lattice vectors
\begin{align}
\label{ps_reciprocalvector}
    \vc{R}_m^{\star} = 2 \pi(m_1\vc{b}_1+m_2\vc{b}_2), \quad m = (m_1,m_2) \in \mathbb{Z}^2.
\end{align}
The fundamental translational vectors $\vc{b_1}$ and $\vc{b_2}$ in $\Lambda^\star$ are related to those in $\Lambda$ through
\begin{equation}
\label{ps_vectorsrel}
   \vc{a}_i\vc{b}_j = \delta_{ij}, \quad i,j=1,2.
\end{equation}

It is shown \cite{Movchan, Linton} that
\begin{align}
\label{ps_rlsum}
    \sigma_n^Y(k_o,\vc{\beta})  = \frac{4 i^n}{A}\sum_{\scriptstyle \vc{R}_m^{\star} \in \Lambda^{\star}}
                                                        \frac{J_n(\beta_m\xi)}{J_n(k_o \xi)}\frac{e^{i n \tau_m}}{k_o^2-\beta_m^2}-
                                                        \delta_{0n} \frac{Y_0(k_o \xi)}{J_0(k_o \xi)},
\end{align}
where $A$ is the area of a lattice cell,$\vc{\beta}_m = \vc{\beta}+\vc{R}_m^{\star} = \beta_m (\cos\tau_m,\sin\tau_m)$  and $\xi \in [0,\zeta]$ with
\begin{align}
\label{ps_zeta}
   \zeta \leqslant \mymin{\vc{R}_j\in\Lambda, \vc{R}_j\neq 0}R_j.
\end{align}

The lattice sum given by \eqref{ps_rlsum} has simple poles $k=\beta_m$ that correspond to a plane wave solutions satisfying Floquet-Bloch conditions \eqref{ps_Bloch} in the absence of the scatterers \cite{McIver1}. The magnitude $\beta_m$ is defined by $M$ pairs of integer numbers $(m_1,m_2)\in\mathbb{Z}_M^2$ that gives $M$ different wave vectors $\vc{\beta}_m$. In the vicinity of this pole the related $M$ terms of the lattice sum \eqref{ps_rlsum} take the leading order that is 
\begin{align}
\label{ps_apxlsum}
		\sigma_n^Y(k_o,\vc{\beta}) = \frac{4 i^n}{A}\sum_{m \in \mathbb{Z}_M^2} 
																									\frac {\exp(i n \tau_m)} {k_o^2-\beta_m^2} + \sigma_n^{Y,2}(k_o,\vc{\beta}), 
\end{align}
where $\sigma_n^{Y,2}(k,\vc{\beta})$ is the next order in expansion over \cite{McIver2}
\begin{align}
\label{ps_orders1}
	(k^2-\beta_m^2)L^2 \ll 1,\;m \in \mathbb{Z}_M^2.
\end{align}

In this section it is assumed that only one pair $(m_1,m_2)=(0,0)$ contributes to the leading order of expansion \eqref{ps_apxlsum}. This results in $\vc{\beta_m}=\vc{\beta}$. It is noted that this is true for the case when $k_o L\ll 1$ and $\beta_m L \ll 1$.

It is also assumed that $\vc{\beta}L$ is taken within the first irreducible Brillouin zone \cite{Nicorovici} that can be described by the contour $(0,0)-(\pi,0)-(\pi,\pi)$.

The smallness of $\vc{\beta}L$ and approximation of lattice sum \eqref{ps_apxlsum} by the summand related to the pair of $(m_1,m_2)=(0,0)$ make valid further approximations along $(0,0)-(\pi,0)$ and $(\pi,\pi)-(0,0)$ lines. This also leads to a symmetry around point $(0,0)$ so that vector $\vc{\beta}$ is only considered within the interval $(0,0)-(\pi,0)$ (i.e. $\beta=q_1$ and $\tau=0$ with $q_1 \in [0,q]$ and $q L \ll \pi$). As a result the leading order of the lattice sum \eqref{ps_apxlsum} is simplified to
\begin{align}
\label{ps_apxlsum0}
		\sigma_n^Y(k,\vc{\beta}) \approx \sigma_n^{Y,1}(k,\vc{\beta}) = \frac{i^n}{A} \frac {4} {k_o^2-\beta^2},
\end{align}

Substituting equations \eqref{ps_apxlsum0} into dispersion relation \eqref{ps_RI} one can obtain the following system of homogeneous algebraic equations with respect to unknown coefficients $B_n^J,\,n \in \mathbb{Z}$
\begin{align}
\label{ps_apxRI}
	B_n^J - \frac{4}{A (k_o^2-\beta^2)} \sum_{p=-\infty}^{+\infty} (-i)^{p-n} B_p^J Z_p =0,\, n \in \mathbb{Z}.
\end{align}
By multiplying both sides of equation \eqref{ps_apxRI} by $(-i)^{n}$ it is possible to redefine the unknown coefficients as $\hat{B}_n^J = (-i)^n B_n^J,\,n \in \mathbb{Z}$ so that the system of equations \eqref{ps_apxRI} becomes similar to that derived in \cite[eq. (71)]{LintonMartin}. This equation can now be employed to derive Foldy's type dispersion relation. Following the derivations in \cite[eqs. (76) and (77)]{LintonMartin} it is concluded that coefficients $\hat{B}_n^J,\,n\in\mathbb{Z}$ are identical for all indices $n\in \mathbb{Z}$, yielding
\begin{equation}
\label{ps_Foldy}
	\beta^2=k_o^2 - \frac{4}{A} F,
\end{equation}
where $\displaystyle{F=\sum_{p=-\infty}^{+\infty} Z_p}$ is determined by boundary conditions on the scatterer's surface. It also coincides with the imaginary part of the far-field pattern of a single scatterer. The consistency of equations \eqref{ps_orders1} and \eqref{ps_Foldy} requires that
\begin{equation}
\label{ps_orders2}
		\frac{4}{A} F L^2 << 1.
\end{equation}
It must be noted that for the resonant scatterers $F$ have singular points that give resonance frequencies. Therefore, in the vicinity of resonances the asymptotic orders of factors $Z_n,\,n\in\mathbb{Z}$ can invalidate the derivation of dispersion relation \eqref{ps_Foldy}. This, however, can also signify the existence of a band gap due to resonance.

Now consider a square periodic array of thin-walled elastic shells with area of lattice cell $A=L^2$. The application of long-wave low-frequency approximations results in the following expressions for factor $Z_n,\,n\in\mathbb{Z}$ described by \cite[eq. (22)]{SUOU}
\begin{align}
\label{ps_shellZ}
		Z_n =-\frac{[J'_n(k_o a)]^2[1-(k_3 a)^2+n^2]}{J'_n(k_o a) Y'_n(k_o a)[1-(k_3 a)^2+n^2]+
    																																			[n^2-(k_3 a)^2]\rho_o(\rho \pi a h)^{-1}},\quad Z_{-n}=Z_n,
\end{align} 
where $\rho$ is the density of an elastic material, $k_3=\omega/c_3$ and $a$ is shell mid-surface radius (referred to as radius of the elastic shell) and $h$ is its half-thickness. Here the dilatational wave speed $c_3$ for a thin elastic plate is recalled
\begin{equation}
\label{ps_c3}
		c_3 = \sqrt{\frac{E}{\rho\left(1-\nu^2\right)}}.
\end{equation}
within which $E$ and $\nu$ are Young's modulus and Poisson ratio of elastic shell, respectively.
 
For the low frequency range that contains $n=0$ (i.e. axisymmetric resonance) and $n=\pm 1$ resonances the far-field pattern of elastic shell can be approximated by its first two terms that gives
\begin{equation}
\label{ps_F01}
		F = Z_0 + 2 Z_1.
\end{equation}
The assumption that wavelength of a propagating acoustic wave is much bigger than the lattice cell size (i.e. lattice constant $L$) immediately results in 
\begin{equation}
\label{ps_smpar}
		\epsilon=k_o a \ll 1. 
\end{equation}
The latter can be used to simplify \eqref{ps_shellZ} by expanding Bessel functions. It is also assumed that the elastic shell is ``soft'' (i.e. $\rho_o/\rho = O(\epsilon^2)$ and $c_3/c = O(\epsilon)$) so that acoustic waves are able to penetrate it. Thus
\begin{align}
\label{ps_Z0}
		Z_0 &\approx \epsilon^2 \frac{\pi}{4} \frac{K_{0}^2 - k_o^2}{\hat{K}_{0}^2-k_o^2},\\
\label{ps_Z1}		
		Z_1 &\approx \epsilon^2 \frac{\pi}{4} \fbrl -1+ \frac{\rho_o a}{\rho h}\frac{K_{0}^2-k_o^2}{2 K_{0}^2-k_o^2} + 
																							\epsilon^2 \left[ \frac 1 2 \log\frac{\epsilon}{2} + \frac 1 8 (5+4\gamma) \right] \fbrr,
\end{align}
where wavenumbers
\begin{align}
\label{ps_K0}
	K_{0}&=\frac{1}{a}\sqrt{\got{K}_c},\\	
\label{ps_K0hat}
	\hat{K}_{0}&=\frac{1}{a}\sqrt{\got{K}_{\rho}+\got{K}_c}
\end{align}
with 
\begin{equation}
\label{ps_Krho}
		\got{K}_{\rho}=\frac{\rho_o}{\rho}\frac{a}{h}
\end{equation}
and
\begin{equation}
\label{ps_Kc}
		\got{K}_{c}=\left(\frac{c_3}{c_o}\right)^2
\end{equation}
correspond to axisymmetric resonances of shell in vacuum with vacuum and acoustic medium inside, respectively. In equation \eqref{ps_Z1} approximated $Z_1$ has a singularity at $k_o^2 = 2 K_0^2$ related to the leading order of resonance frequency with index $n=1$. From this approximation it can be concluded that frequency of $n=1$ resonance of a thin elastic shell  defined from
\begin{equation}
\label{ps_K1}
		K_1= \sqrt{2} K_0 
\end{equation}
has no dependence on the physical parameters of the surrounding acoustic environment in the leading order.

\begin{figure}[t]
		\center
		\subfigure[]{\includegraphics[scale=1]{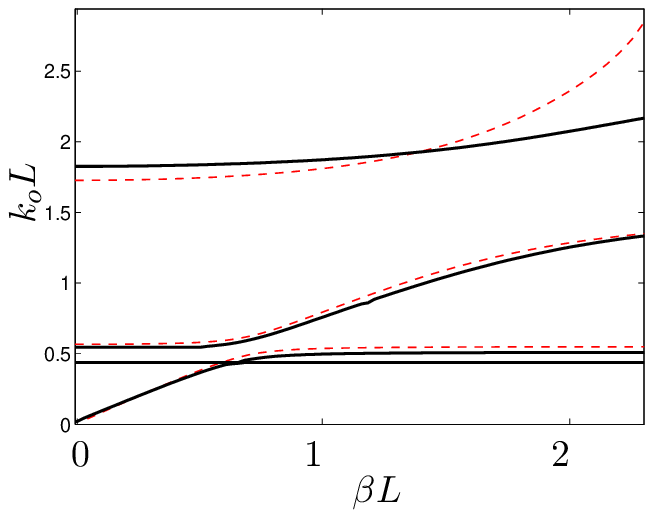}}\hspace{2mm}
		\subfigure[]{\includegraphics[scale=1]{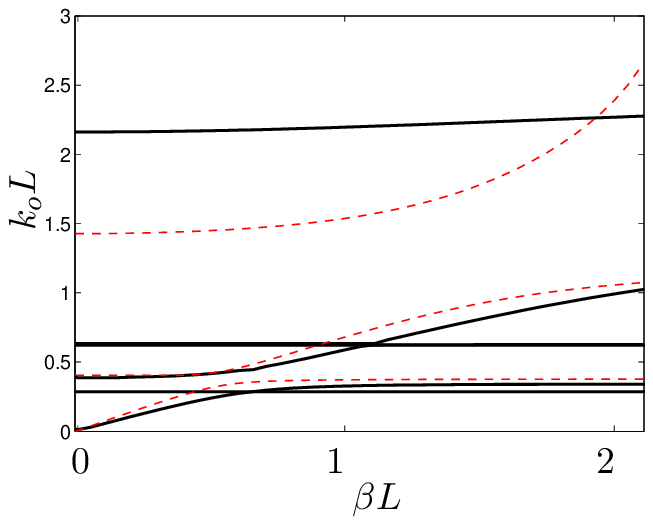}}
    \caption{(Color online) Foldy's approximation \eqref{ps_Foldy} ('--~--') compared with the semi-analytical solution of Rayleigh Identity \eqref{ps_RI} (solid line). (a) $a=0.0275$ m and (c) $a=0.0375$ m.}
    \label{fig:Foldy}
\end{figure}

As $\beta L \rightarrow 0$ equation \eqref{ps_Foldy} gives two non-zero solutions for $k_o$ in addition to $k_o=0$. These solutions can be used to find the size of the corresponding band gaps.

In the vicinity of axisymmetric resonance $\hat{K}_0$ as $\beta L \rightarrow 0$ the dispersion relation \eqref{ps_Foldy} can be transformed to
\begin{equation}
\label{ps_apxFoldyK0}	
	\got{F} \got{K}_{\rho} (\got{K}_{c}-\got{K}_{\rho}) + \left[(k_o a)^2 -
	 					(\got{K}_{c}+\got{K}_{\rho})\right]\left[(1+\got{F})(\got{K}_{\rho}-\got{K}_{c}) - \got{F}\got{K}_{\rho}(1+2\got{K}_{\rho})\right]=0.
\end{equation}
This results in the following solution for $k_o > 0$
\begin{align}
\label{ps_uplimK0}
		k_{o,2} &= \frac{1}{a}\sqrt{\got{K}_{\rho}+\got{K}_{c} + \frac{\got{F}\got{K}_{\rho}(\got{K}_{\rho}-\got{K}_{c})}{\got{K}_{\rho}-\got{K}_{c}- \got{F}(\got{K}_{c}+2\got{K}_{\rho}^2)}} \approx \frac{1}{a}\sqrt{a^2 \hat{K}_{0}^2 + \got{F} \got{K}_{\rho}}
\end{align}
where $\got{F}=\pi a^2/A$ is the scatterer filling fraction. Expression \eqref{ps_uplimK0} immediately estimates the upper limit of the band gap. Note that the lower limit of the mentioned band gap is given by $k_{o,1} = \hat{K}_0 L$. Therefore, based on the derived approximations, it is possible to conclude that in the doubly periodic array of thin elastic shells (with filling fraction $\got{F}\ll 1$) waves do not propagate in the frequency interval 
\begin{equation}
\label{ps_n0bandgap}
	\left[f^l_{n=0}, f^u_{n=0}\right]=\left[\frac{c_3}{2\pi a} \sqrt{1+\frac{\got{K}_{\rho}}{\got{K}_c}}, 
	\frac{c_3}{2\pi a}\sqrt{1+\frac{\got{K}_{\rho}}{\got{K}_{c}}(1+\got{F})}\right].
\end{equation}

In the vicinity of $n=1$ resonance $K_1$ is approximated by \eqref{ps_K1} and as $\beta L \rightarrow 0$ the dispersion relation \eqref{ps_Foldy} can be transformed to
\begin{equation}
\label{ps_apxFoldyK1}	
	2 \got{F} \got{K}_{\rho} \got{K}_{c} (\got{K}_{\rho}-\got{K}_{c}) + \left[(k_o a)^2 -
	 					2\got{K}_{c})\right]\left[(1+2\got{F}-2\got{F}\got{K_{\rho}})(\got{K}_{c}-\got{K}_{\rho}) -
	 					 \got{F}\got{K}_{c}(1+2\got{K}_{\rho})\right]=0.
\end{equation}
For $k_o > 0$ the solution of this equation takes the following form
\begin{align}
\label{ps_uplimK1}
		k_{o,2} &= \frac{1}{a}\sqrt{ 2 \got{K}_{c} + \frac{2 \got{F}\got{K}_{\rho}\got{K}_{c}(\got{K}_{c}-\got{K}_{\rho})}
		{\got{K}_{c}-\got{K}_{\rho}+\got{F}\left[2(\got{K}_{c}-\got{K}_{\rho})(1-\got{K}_{\rho})-\got{K}_{c}(1+2\got{K}_{\rho})\right]}}\nonumber\\
		&\approx \frac{1}{a}\sqrt{a^2 K_{1}^2 + 2 \got{F} \got{K}_{\rho} \got{K}_{c}}
\end{align}
The lower limit of the band gap related to $n=1$ resonance is given by equation \eqref{ps_K1}. Therefore the frequency interval where waves do not propagate is estimated as
\begin{equation}
\label{ps_n1bandgap}
	\left[f^l_{n=1}, f^u_{n=1}\right]=\left[\frac{\sqrt{2}c_3}{2\pi a}, 
	\frac{\sqrt{2}c_3}{2\pi a}\sqrt{1+\got{F}\got{K}_{\rho}}\right].
\end{equation}

In Figure \ref{fig:Foldy} the solutions of dispersion relation \eqref{ps_Foldy} obtained for $F$ that contains only two terms $n=0,1$ are shown. Throughout this paper unless otherwise specified the material parameters of thin elastic shell are identical to those in \cite{SUOU} and its thickness $2h=0.00025$ m. The radius of the elastic shell is varied whereas the lattice constant $L$ is fixed ($L=0.08$ m) so that two values of the filling fraction $\got{F}$ are considered: 0.4 and 0.6. Two band gaps related to the resonances $\hat{K}_0$, $n=0$ and $\sqrt{2}K_0$, $n=1$ are observed below the first band gap associated with the array periodicity. According to Figure~\ref{fig:Foldy}(a) the estimates of the lower and upper limits of the band gap are within $5\symbol{37}$ of the exact values as long as the filling fraction is relatively low ($\got{F} \geq 0.4$). However as the filling fraction is increased up to $0.6$ the accuracy is expected to decrease according to the condition \eqref{ps_orders2}. This is illustrated in Figure~\ref{fig:Foldy}(b) where estimated value of the upper boundary of the band gap associated with $n = 0$ is $30\symbol{37}$ in error from the exact value. It must also be noted that this upper boundary obtained with Foldy's equation depends on the form of the factor $Z_n$. Its approximated form might give more accurate results than the original form of $Z_n$ given by \eqref{ps_shellZ}. This inconsistency points out that to obtain more accurate and consistent results equation \eqref{ps_Foldy} has to be modified to include the periodicity effects.

\section{Matched asymptotic expansion}\label{MAE}

The results obtained in the previous section can be improved by using the technique based on matched asymptotic expansions (MAE) \cite{McIver1,Crighton}. This technique allows approximating eigenvalues of \eqref{ps_Helmholtz}-\eqref{ps_Bloch} when $k_o L = O(1)$ in the vicinity of band gaps associated with the periodicity (Bragg band gaps). However in this paper MAE is used to estimate the solution that forms the upper boundary of the band gap due to the axisymmetric resonance ($n=0$) of a thin elastic shell.

First outer and inner regions are introduced. In the inner region surrounding each scatterer the solution is solved in conjunction with the boundary conditions imposed on the surface of the scatterer. The characteristic length of this region is the radius $a$ of scatterer. Thus a dimensionless inner coordinate is given by  
\begin{equation}
\label{mae_incoord}
		x=\frac{r}{a}.
\end{equation}
In the outer region the scatterers are replaced by point sources and the solution is subject to quasi-periodicity conditions. The characteristic length in this region is equal to the wavelength of sound. Thus a dimensionless outer coordinate is given by
\begin{equation}
\label{mae_outcoord}
		y=k_o r.
\end{equation}
The wavelength is assumed to be much bigger than the scatterer radius. This requirement results in small parameter $\epsilon$ introduced in Section \ref{Foldy} that connects two regions through
\begin{equation}
\label{mae_ge}
		y=\epsilon x.
\end{equation}

\subsection{Inner solution}

The inner solution is found from the boundary value problem for a single thin elastic shell. The solution around the cylinder is represented as
\begin{equation}
\label{mae_innersol}
    \psi(x,\theta) = \sum_{n=-\infty}^{+\infty} B_n^J\left[ J_n(\epsilon x) + Z_n Y_n(\epsilon x)\right] e^{i n\theta}
\end{equation}
with unknown constants $B_n^J$ and coefficients $Z_n$ defined by equation \eqref{ps_shellZ}. According to equation \eqref{ps_Z0}, factor $Z_0$ takes order $O(\epsilon^2)$ if its singular points (i.e. resonances) are isolated. In the vicinity of axisymmetric resonance the order of factor $Z_0$ can be transformed to $O(\eta)$ that is
\begin{equation}
\label{mae_Z0}
		Z_0 = \delta_Z \eta,\quad \textrm{with}\quad \delta_Z = O(1),
\end{equation}
where small parameter $\eta$ is bigger than $\epsilon$ and is assumed to be
\begin{equation}
\label{mae_eta}
		\eta=\frac{1}{K-\log{\epsilon}},
\end{equation}
within which $K$ is the unknown constant.

By assuming $Z_0$ of order $\eta$ the proximity of $k_0$ to the axisymmetric resonance $\hat{K}_0$ is restricted by
\begin{equation}
\label{mae_koK0}
		k_o-\hat{K}_0 = O \left(\frac{\epsilon^2}{\eta}\right)
\end{equation}

Expanding solution \eqref{mae_innersol} up to the order $\eta$ first with respect to the inner coordinate $x$ and then with respect to the outer coordinate $y$ results in
\begin{equation}
\label{mae_innersolexp}
		\psi^{(\eta,\eta)}(x,\theta) = B_0^J \left[1 - \frac{2}{\pi} \delta_Z + \eta \frac{2}{\pi} \delta_Z \log{x}\right].
\end{equation}
It is noted that the terms in $O(1)$ and $O(\eta)$ orders have no azimuthal dependence.

\subsection{Outer solution}
The outer solution of the Helmholtz equation \eqref{ps_Helmholtz} satisfies the quasi-periodicity conditions \eqref{ps_Bloch} and is singular at the points $O_j$. This results in
\begin{equation}
\label{mae_source}
    \Psi(r,\theta) = \sum_{n=-\infty}^{+\infty}A_n\sum_{\vc{R}_j \in \Lambda\setminus\{\vc{0}\}}e^{i\vc{\beta}\vc{R}_j}\Hn{n}{1}(k_o r_j)e^{i n \theta_j},
\end{equation}
where local coordinates $(r_j,\theta_j)$ with origin at $O_j$ are used. 

It is assumed that the leading order of coefficients $A_n,\,n\in\mathbb{Z}$ is bounded by the small parameter $\eta$. This approach is similar to that derived in \cite{McIver2} and \cite[Section 6.8]{Crighton}. Thus
\begin{equation}
\label{mae_outerscaling}
		A_n = \eta \hat{A}_n.
\end{equation}

The use of addition theorem for the Bessel functions \cite{Abramowitz} and lattice sum representation in the reciprocal lattice $\Lambda^\star$ transform the outer solution \eqref{mae_source} to
\begin{equation}
\label{mae_outersol}
		\Psi^{(\eta)}(y,\theta) = \eta \sum_{n=-\infty}^{+\infty} \hat{A}_n \fbrl  \Hn{n}{1}(y)e^{i n\theta} + \sum_{p=-\infty}^{+\infty}(-1)^{n-p}
                                                                														\sigma_{n-p}(k_o,\vc{\beta})J_p(y)e^{i p \theta}\fbrr
\end{equation}
and
\begin{equation}
\label{mae_lsum}
		\sigma_{n}(k_o,\vc{\beta}) = -\delta_{0,n} + i \sigma_{n}^Y(k_o,\vc{\beta}),
\end{equation}
where $\sigma_{n}^Y(k_o,\vc{\beta})$ is given by equation \eqref{ps_rlsum}.

The form of the inner solution \eqref{mae_innersolexp} assumes that the leading order should include axisymmetric terms only. In contrast to Section~\ref{Foldy} it is assumed here that for $n=0$ the entire lattice sum \eqref{ps_rlsum} (not only pair $(m_1,m_2)=(0,0)$) contributes to the outer solution as
\begin{equation}
\label{mae_kb}
		\sigma_{0}^Y(k_o,\vc{\beta}) = \frac{\delta}{\eta},\quad\textrm{with}\quad \delta=O(1).
\end{equation}
Then expanding outer solution in terms of the inner coordinate $x$ up to order $\eta$ leads to
\begin{equation}
\label{mae_outersolexp}
		\Psi^{(\eta,\eta)}(x,\theta) = i \hat{A}_0 \left[\delta - \frac{2}{\pi} + 
												\eta \frac{2}{\pi} \left( K + \gamma - \log{2} + \log{x}\right)\right],
\end{equation}

\begin{figure}[t]
		\center
		\subfigure[]{\includegraphics[scale=1]{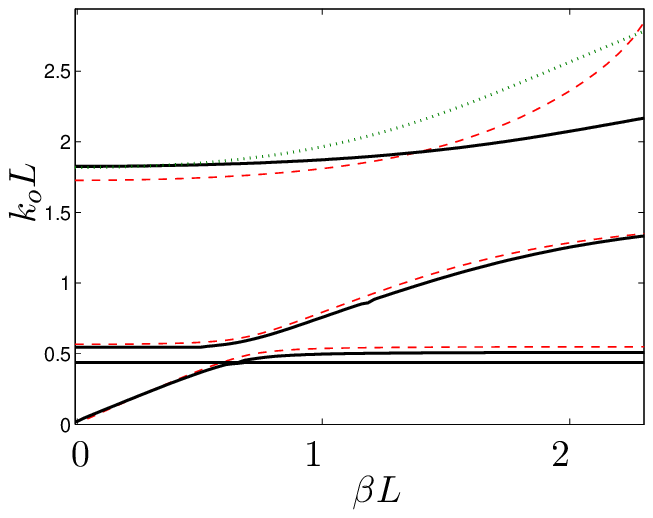}}\hspace{2mm}
		\subfigure[]{\includegraphics[scale=1]{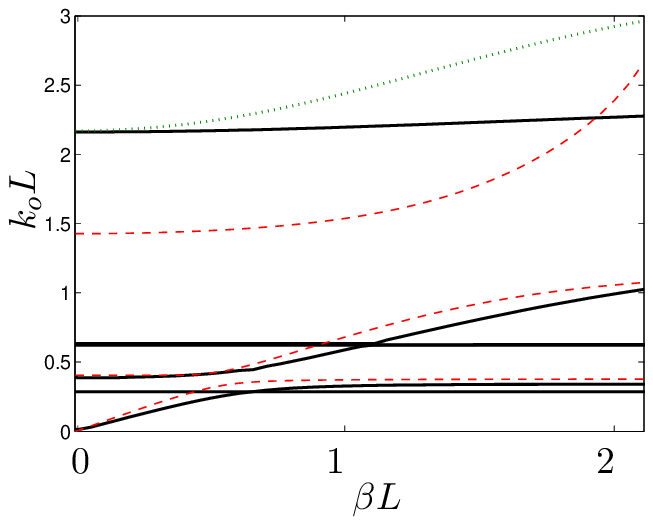}}
    \caption{(Color online) MAE approximation of the upper limit \eqref{mae_ulim} ('$\cdot\cdot\cdot$') compared with the Foldy's approximation \eqref{ps_Foldy} ('--~--') and the semi-analytical solution of Rayleigh Identity \eqref{ps_RI} (solid line). (a) $a=0.0275$ and (b) $a=0.0375$.}
    \label{fig:MAE}
\end{figure}

\subsection{Matching}

The inner expansion of the outer solution \eqref{mae_outersolexp} can now be matched with the outer expansion of the inner solution \eqref{mae_innersolexp}. First factors of $\log{\xi}$ in order $\eta$ give
\begin{equation}
\label{mae_A0B0}
		\hat{A}_0 = -i \delta_Z B_0^J.
\end{equation}
Matching constants in order $\eta$ results in
\begin{equation}
\label{mae_K}
		K = \log{2} - \gamma.
\end{equation}
Finally the consistency of the leading orders in \eqref{mae_outersolexp} and \eqref{mae_innersolexp} requires that
\begin{equation}
\label{mae_disprel}
		Z_0 \sigma_0^{Y}(k_o,\beta) - 1 = 0,
\end{equation}
where equations \eqref{mae_Z0}, \eqref{mae_kb} and \eqref{mae_A0B0} have been used.

As $\beta L \rightarrow 0$ the lattice sum in equation \eqref{mae_disprel} can be approximated by
\begin{equation}
\label{mae_lsumexp}
		\sigma_0^{Y}(k_o,\beta L \rightarrow 0) \approx \frac{1}{J_0(k_o \xi)} \left[\frac{4 L^2}{A} \frac{1}{(k_o L)^2}
		-Y_0(k_o \xi)
		+\frac{4 L^2}{A}
		\sum_{\overset {\scriptstyle \vc{R}_m^{\star} \in \Lambda^{\star}} {\scriptstyle \vc{R}_m^{\star}\neq(0,0)}}
		\frac{J_0(R_m^{\star} \xi)}{(k_oL)^2-(R_m^{\star} L)^2}\right].
\end{equation}
It is noted that the last term given as a series over the reciprocal lattice represents the correction factor to the upper limit of the band gap related to the axisymmetric resonance. This correction factor includes the periodicity effects.

For the square lattice with lattice constant $L$ and $\xi=L/2$ the lattice sum \eqref{mae_lsumexp} is transformed into
\begin{align}
\label{mae_lsumexpL}
		\sigma_0^{Y}(k_o,\beta L=0) = \frac{1}{J_0(k_o L/2)}\left[\frac{4}{(k_o L)^2}-Y_0(k_o L/2) + 4 S(k_o L)\right],
\end{align}
with
\begin{align}
\label{mae_lsumS}
	S(k_o L)=\sum_{\overset {\scriptstyle m \in \mathbb{Z}^2} {m\neq(0,0)}}\frac{J_0\left(\pi\sqrt{m_1^2+m_2^2}\right)}{(k_o L)^2-4\pi^2(m_1^2+m_2^2)}.
\end{align}
It is noted that $\xi$ is chosen from the interval set in \eqref{ps_zeta} to improve the convergence of $S$.

The lattice sum \eqref{mae_lsumexpL} and the dispersion relation \eqref{mae_disprel} can be used to improve the upper limit of the band gap found in \eqref{ps_uplimK0}. Thus the improved estimate is given by the solution of the following equation
\begin{align}
\label{mae_ulim}
	Z_0\left\{4 - (k_o L)^2\left[Y_0(k_o L)+4 S(k_o L)\right]\right\} - (k_o L)^2 J_0(k_o L) = 0,
\end{align}
within which $k_o L \neq 0$ and $J_0(k_o L)\neq 0$. To find the solution of equation \eqref{mae_ulim} the infinite sum \eqref{mae_lsumS} has to be truncated. The results are accurate to three decimal places for both $|m_1| \geq 8$ and $|m_2| \geq 8$.

Figure \ref{fig:MAE} demonstrates the improvement in approximation of the upper boundary of the band gap due to axisymmetric resonance given by equation \eqref{mae_ulim} compared to Foldy's approximation equation \eqref{ps_uplimK0}. For a low filling fraction (i.e. $\got{F}<0.4$) the results are accurate within $1\symbol{37}$ from those obtained with the Foldy's equation, see Figure~\ref{fig:MAE}(a). As the filling fraction increases the approximation of the branch representing the upper bound deteriorates. However the limiting point at $\beta L=0$ and its vicinity are not affected, see Figure~\ref{fig:MAE}(b) so that the estimate remains valid even for large filling fraction ($\got{F}>0.4$) with the accuracy within $1\symbol{37}$ from the exact solution.

\section{Self-consistent effective medium formulation}\label{CPA}

The effective medium approach provides an alternative method for estimating the boundaries of band gaps due to resonances with indices $n=0$ and $n=1$.

\begin{figure}[t]
		\center
		\includegraphics[scale=1]{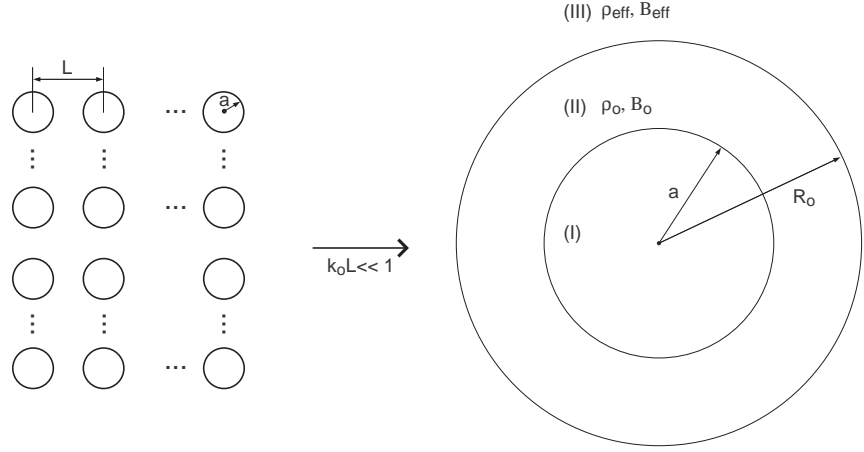}
    \caption{Geometry of the effective medium}
    \label{fig:1layergeom}
\end{figure}

Consider an effective medium as a replacement of a periodic array of scatterers in a fluid matrix (see Figure~\ref{fig:1layergeom}). The homogenization \cite{Sheng,Norris} becomes possible only when the wavelength in both fluid and effective medium is much bigger than the radius of scatterers $a$ and separation distance between them (lattice constant $L$) that is $k_o L \ll 1$ and $k_{eff} L \ll 1$.

The solution of the Helmholtz equation
\begin{equation}
\label{eff_Helmholtz}
	\Delta p_\alpha + k_\alpha^2 p_\alpha=0,
\end{equation}
is given in terms of the displacement potential $p_\alpha(\vc{r})$ where $k_\alpha=\omega/c_\alpha$ is the wavenumber, $c_\alpha=\sqrt{B_\alpha/\rho_\alpha}$ is speed of sound, index $\alpha$ relates solution $p$ to one of the regions (i.e. '$eff$' effective medium (I), '$o$' matching fluid layer between scatterer and effective medium). The problem is solved in polar coordinates $\vc{r}=(r,\theta)$. It is also noted that the physical parameters of the matching fluid layer coincide with the acoustic medium introduced in Section~\ref{Foldy}.

To find the effective medium parameters (effective density $\rho_{eff}$ and bulk modulus $B_{eff}$) and hence its wavenumber $k_{eff}=\omega\sqrt{\rho_{eff}/B_{eff}}$ the problem is split into three regions that are \cite{Mei}
\begin{mylist3}
\item - circular scatterer,
\item - fluid layer with properties identical to those of the fluid matrix,
\item - effective medium with yet unknown properties,
\end{mylist3}
where regions (I) and (II) are introduced to derive parameters of the effective medium through matching acoustic potentials at the interface of the region (II) and (III) $r=R_o$. That is
\begin{align}
\label{eff_slitbc}
	p_o 									&= \frac{\rho_{eff}}{\rho_o} p_{eff},\nonumber\\
	\frac{\pd p_o}{\pd r}	&= \frac{\pd p_{eff}}{\pd r},
\end{align}
where outer radius $R_o$ is derived in terms of the scatterer radius and the filling fraction  of the original periodic array with lattice constant $L$ so that the filling fraction in a composite inclusion remains equal to that of the original periodic array. This gives \cite{Yang}
\begin{equation}
\label{eff_Ro}
	R_o=\frac{a}{\sqrt{\got{F}}}.
\end{equation}

The total wave field in region (III) is given by
\begin{align}
\label{eff_efftsol}
    p_{eff}&=p_{inc}+p_{scat},\\
\label{eff_effincsol}
		p_{inc}&=e^{i k_{eff} r cos(\theta-\beta)},
\end{align}
where $p_{inc}$ is plane wave incident at an angle $\beta$ and $p_{scat}$ represents scattered wave field. Having assumed that the effective medium behaves as a homogeneous fluid the scattered wave field vanishes that leads to
\begin{align}
\label{eff_efftmsol}
    p_{eff}=p_{inc},
\end{align}
with the plane wave expanded over the regular Bessel functions $J_n(k_{eff}r)$ as \cite[eqs. 9.1.44 and 9.1.45]{Abramowitz}
\begin{align}
\label{eff_pwBessel}
    p_{inc}=\sum_{n=-\infty}^{+\infty} i^n J_n(k_{eff}r) e^{i n (\theta-\beta)},
\end{align}

Solution inside the annular layer (II) takes the following form
\begin{align}
\label{eff_layersol}
    p_o=\sum_{n=-\infty}^{+\infty}B_n\left[J_n(k_o r) + Z_n Y_n(k_o r)\right]e^{i n \theta},
\end{align}
where factors $Z_n$ are defined by \eqref{ps_shellZ}.

Substitution of equations \eqref{eff_pwBessel}and \eqref{eff_layersol} into the boundary conditions \eqref{eff_slitbc} leads to
\begin{align}
\label{eff_sys}
		B_n\left[J_n(k_o R_o) + Z_n Y_n(k_o R_o)\right] 	&= \frac{\rho_{eff}}{\rho_o} J_n(k_{eff} R_o) e^{i n (\pi/2-\beta)},\nonumber\\
		B_n\left[J'_n(k_o R_o) + Z_n Y'_n(k_o R_o)\right] &= J'_n(k_{eff} R_o) e^{i n (\pi/2-\beta)},
\end{align}
where $n\in\mathbb{Z}$. To solve system \eqref{eff_sys} it first has to be truncated at some integer number $N$ that gives
\begin{align}
\label{eff_truncsys}
\vc{A} \vc{x}=\vc{b}
\end{align}
where matrix $\vc{A}$ has $4N$ rows and $2N$ columns, vector $\vc{x}$ has $2N$ elements and vector $\vc{b}$ has $4N$ elements, and
\begin{align}
\label{eff_A}
\vc{A}&=
\left(\begin{array}{cccc}
		\left[J_{-N}(k_o R_o) + Z_{-N} Y_{-N}(k_o R_o)\right] 		& 0 & \ldots & 0\\
		\left[J'_{-N}(k_o R_o) + Z_{-N} Y'_{-N}(k_o R_o)\right] 	& 0 & \ldots & 0\\
		\vdots & \vdots &  & \vdots\\
		0	& 0 & \ldots & \left[J_{N}(k_o R_o) + Z_{N} Y_{N}(k_o R_o)\right]\\
		0 & 0 & \ldots & \left[J'_{N}(k_o R_o) + Z_{N} Y'_{N}(k_o R_o)\right]\\		
\end{array}\right),\\
\label{eff_x}
\vc{x}&=
\left(\begin{array}{l}
	B_{-N}\\
	B_{-N+1}\\
	\vdots\\
	B_{N}
\end{array}\right),\\
\label{eff_b}
\vc{b}&=
\left(\begin{array}{l}
	\displaystyle \frac{\rho_{eff}}{\rho_o} J_{-N}(k_{eff} R_o) e^{- i N (\pi/2-\beta)}\\
	\displaystyle J'_{-N}(k_{eff} R_o) e^{- i N (\pi/2-\beta)}\\	
	\vdots\\
	\displaystyle \frac{\rho_{eff}}{\rho_o} J_{N}(k_{eff} R_o) e^{i N (\pi/2-\beta)}\\
	\displaystyle J'_{N}(k_{eff} R_o) e^{i N (\pi/2-\beta)}\\
\end{array}\right).
\end{align}

According to Kronecker-Capelli theorem \cite{Shilov} the overdetermined system \eqref{eff_truncsys} is compatible if the rank of its coefficient matrix is equal to that of the augmented matrix. On the other hand the unique solution of system \eqref{eff_truncsys} exists if the rank of matrix $\vc{A}$ is equal to number of variables (i.e. number of columns $2N$). Hence combination of these two statements and application of Gaussian elimination algorithm to the augmented matrix $(\vc{A}|\vc{b})$ yield the criteria for the existence of the solution that is
\begin{align}
\label{eff_syssol}
		\frac{J'_n(k_o R_o) + Z_n Y'_n(k_o R_o)}{J_n(k_o R_o) + Z_n Y_n(k_o R_o)} 			
																		&= \frac{\rho_o}{\rho_{eff}} \frac{J'_n(k_{eff} R_o)}{J_n(k_{eff} R_o)},\quad n=-N..N.
\end{align}
These equations can now be simplified by recalling the assumptions 
\begin{equation}
\label{eff_eta}
	\eta=k_o R_o \ll 1\quad \textrm{and}\quad k_{eff} R_o \ll 1.
\end{equation}
Also equation \eqref{eff_syssol} needs to be rewritten as
\begin{align}
\label{eff_syssoldls}
		\frac{\dot{J}_n(\eta) + Z_n \dot{Y}_n(\eta)}{J_n(\eta) + Z_n Y_n(\eta)} 			
																		&= \xi_{\rho}\frac{\dot{J}_n(\eta\xi_{c})}{J_n(\eta\xi_{c})},\quad n=-N..N,
\end{align}
where '${\cdot}$' stands for the derivative of Bessel functions with respect to dimensionless parameter $\eta$. Parameters $\xi_{\rho}$ and $\xi_c$ depend on the effective medium properties as
\begin{align}
\label{eff_xirho}
		\xi_{\rho}&=\frac{\rho_o}{\rho_{eff}},\\
\label{eff_xic}		
		\xi_{c}&=\frac{c_o}{c_{eff}}.
\end{align}
Bessel functions and their derivatives in \eqref{eff_syssoldls} can be replaced by their approximations in the leading orders as $\eta \rightarrow 0$ \cite{Abramowitz}, yielding
\begin{align}
\label{eff_BesselapxJ}
		J_n(\eta)&\approx \left(\frac \eta 2 \right)^n\frac{1}{n!},\; n \geq 0,\\
\label{eff_BesselapxY}
		Y_n(\eta)&\approx -\left(\frac \eta 2 \right)^{-n}\frac{(n-1)!}{\pi},\; n > 0,\\
\label{eff_BesselapxdJ}
		\dot{J}_n(\eta)&\approx \left(\frac \eta 2 \right)^n\left[\frac{n}{n!}\eta^{-1}-\frac{1}{2(n+1)!}\eta\right],\; n \geq 0,\\
\label{eff_BesselapxdY}
		\dot{Y}_n(\eta)&\approx \left(\frac \eta 2 \right)^{-n-1}\frac{n!(1+\delta_{0,n})}{2\pi},\; n \geq 0.
\end{align}
It is convenient to consider subsequently that Bessel functions have only zero and positive integer orders with $Y_0(\eta)=2/\pi\log(\eta)$. For the negative orders of Bessel functions $f_n(\eta)=\{J_{n}(\eta),Y_{n}(\eta)\}$ equation \eqref{eff_syssoldls} is identical to that with $n>0$ due to the relation $f_{-n}(\eta)=(-1)^n f_{n}(\eta)$.

Substitution of equations \eqref{eff_BesselapxJ}-\eqref{eff_BesselapxdY} into equations \eqref{eff_syssoldls} gives for $n=0$
\begin{align}
\label{eff_syssoldlsapx0}
		\frac{(\eta/2)^2-Z_0/\pi}{1 + 2 Z_0 \log(\eta)/\pi} = \xi_{\rho} \xi_{c}^2 \left(\frac \eta 2 \right)^2
\end{align}
and for $n>0$
\begin{align}
\label{eff_syssoldlsapxN}
		\frac{(\eta/2)^{2n}[(n-1)!]^{-1}\left\{1/2-(\eta/2)^2[n(n+1)]^{-1}\right\} + Z_n n! (2\pi)^{-1}}
				 {(\eta/2)^{2n}(n!)^{-1} - Z_n (n-1)! \pi^{-1}}=
		\frac{\xi_{\rho}}{2}\left[n-\left(\frac{\eta\xi_{c}}{2}\right)^2\frac{2}{n+1}\right]
\end{align}
Following the orders of smallness involved in equations \eqref{eff_syssoldlsapx0} and \eqref{eff_syssoldlsapxN} the factor $Z_n$ has to appear as
\begin{equation}
\label{eff_factorC}
		Z_n = \left\{\begin{array}{ll}
								\displaystyle{(\eta/2)^2 \pi Z_{0,0}} & \textrm{for}\quad n=0,\\
								\displaystyle{(\eta/2)^{2n} \frac{\pi}{(n-1)! n!} Z_{n,0}} & \textrm{for}\quad n>0.
								\end{array}\right.
\end{equation}
This assumption can be easily justified for the circular scatterers by expanding $Z_n$ with respect to the small parameter, see \eqref{ps_Z0} and \eqref{ps_Z1}.

By collecting the same orders of smallness in \eqref{eff_syssoldlsapx0} and \eqref{eff_syssoldlsapxN} the leading orders can be derived in the following form
\begin{align}
\label{eff_syssoldlsapx00}
		&\xi_{\rho} \xi_{c}^2 = 1-Z_{0,0},\; n=0,\\
\label{eff_syssoldlsapxN0}		
		&\xi_{\rho} = \frac{1+Z_{n,0}}{1-Z_{n,0}},\; n>0.
\end{align}
From these two equations the parameters of the effective medium are derived in the following order
\begin{align}
\label{eff_B}
		&B_{eff}=B_o\frac{1}{1-Z_{0,0}},\; n=0,\\
\label{eff_rho}		
		&\rho_{eff}=\rho_{o}\frac{1-Z_{n,0}}{1+Z_{n,0}},\;n>0,
\end{align}
where  $B_o=c_o^2 \rho_o$ is bulk modulus of matching fluid layer (i.e. air).

The effective density \eqref{eff_rho} depends on index $n$ since coefficient $Z_{n,0} \sim \got{F}^n$. This limits the use of the proposed model and can only include the contribution of two harmonics that are $n=0$ and any non-zero index $n$. The usual choice of $n=0$ and $n=1$ which are the harmonics that contribute most is followed now. Using \eqref{ps_Z0} and \eqref{ps_Z1} the following expressions are derived for effective density and bulk modulus:
\begin{align}
\label{eff_B0}
	\frac{B_{eff}}{B_o}&=\frac{\got{K}_c+\got{K}_{\rho}-(k_o a)^2}{\got{K}_c(1-\got{F})+\got{K}_{\rho}-(k_o a)^2(1-\got{F})},\\
\label{eff_rho1}
	\frac{\rho_{eff}}{\rho_o}&=\frac{\got{K}_c[2+\got{F}(2-\got{K}_{\rho})]-(k_o a)^2[1+\got{F}(1-\got{K}_{\rho})]}
																	{\got{K}_c[2-\got{F}(2-\got{K}_{\rho})]-(k_o a)^2[1-\got{F}(1-\got{K}_{\rho})]},
\end{align}
where $\got{K}_{\rho}$ and $\got{K}_c$ are defined by \eqref{ps_Krho} and \eqref{ps_Kc}.

\begin{figure}[t]
		\center
		\subfigure[]{\includegraphics[scale=1]{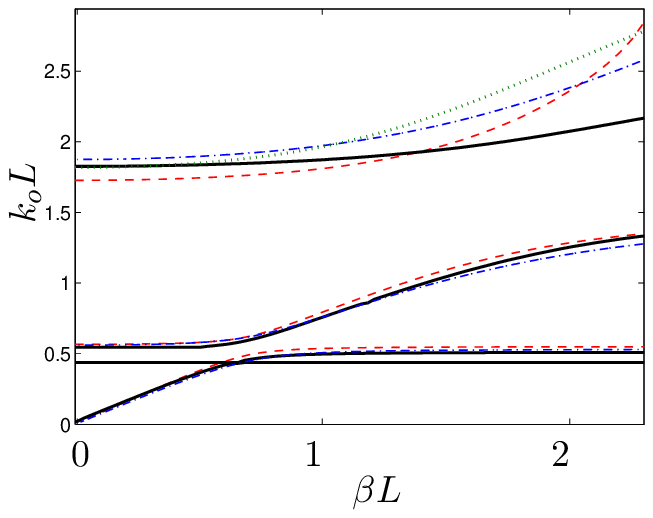}}\hspace{2mm}
		\subfigure[]{\includegraphics[scale=1]{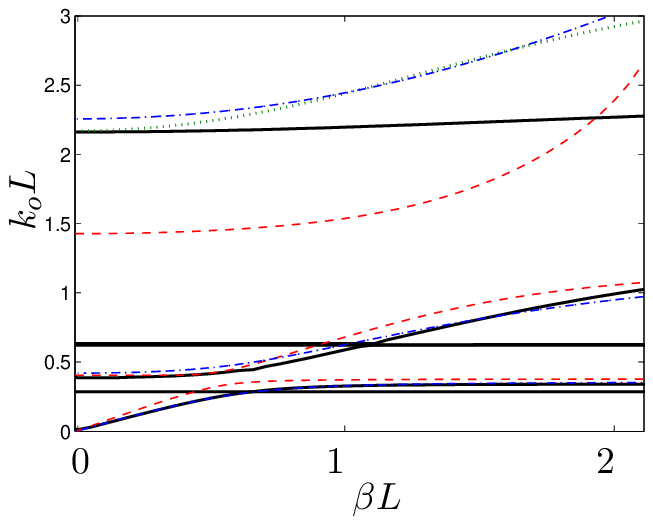}}
    \caption{(Color online) Dispersion relation predicted by self-consistent method \eqref{eff_disprel} ('--~$\cdot$~--') compared with the semi-analytical solution of Rayleigh Identity \eqref{ps_RI} (solid line), Foldy's approximations \eqref{ps_Foldy} ('--~--') and MAE solution \eqref{mae_ulim} ('$\cdot\cdot\cdot$'). (a) $a=0.0275$ and (b) $a=0.0375$.}
    \label{fig:CPA}
\end{figure}

The dispersion relation of the effective medium is
\begin{align}
\label{eff_disprel}
	k_{eff}^2=\frac{\rho_{eff}}{\rho_o }\frac{B_o}{B_{eff}}k_o^2,
\end{align}
which has two poles corresponding to resonances with $n=0$ and $n=1$. Moreover there are two intervals where $k_{eff}$ is imaginary, i.e. band gaps. In the first interval the bulk modulus $B_{eff}$ in \eqref{eff_B0} is negative,
\begin{align}
	\label{eff_lim0l}
	k_o &>\frac{1}{a}\sqrt{\got{K}_{\rho}+\got{K}_c}=\hat{K}_0,\\
	\label{eff_lim0u}
	k_o &<\frac{1}{a}\sqrt{\got{K}_{\rho}+\got{K}_c+\frac{\got{F}\got{K}_{\rho}}{1-\got{F}}}
																				\approxF \frac{1}{a}\sqrt{a^2\hat{K}_0^2+\got{F}\got{K}_{\rho}},
\end{align}
whereas the second interval gives negative effective density $\rho_{eff}$,
\begin{align}																				
	\label{eff_lim1l}
	k_o &>\frac{1}{a}\sqrt{2\got{K}_{c}-\frac{\got{F}\got{K}_{\rho}\got{K}_c}{1-\got{F}(1-\got{K}_{\rho})}}
																										\approxF\frac{1}{a}\sqrt{a^2\hat{K}_1^2-\got{F}\got{K}_{\rho}\got{K}_c},\\
	\label{eff_lim1u}																										
	k_o &<\frac{1}{a}\sqrt{2\got{K}_{c}+\frac{\got{F}\got{K}_{\rho}\got{K}_c}{1+\got{F}(1-\got{K}_{\rho})}}
																										\approxF\frac{1}{a}\sqrt{a^2\hat{K}_1^2+\got{F}\got{K}_{\rho}\got{K}_c}.
\end{align}
Therefore the band gap due to the axisymmetric resonance corresponds to the frequency range where effective bulk modulus is negative:
\begin{equation}
\label{eff_n0bandgap}
	\left[\hat{f}^l_{n=0}, \hat{f}^u_{n=0}\right] =
	\left[\frac{c_3}{2\pi a} \sqrt{1+\frac{\got{K_{\rho}}}{\got{K}_{c}}}, 
			  \frac{c_3}{2\pi a}\sqrt{1+\frac{\got{K}_{\rho}}{\got{K}_c}\left(1+\frac{\got{F}}{1-\got{F}}\right)}\right]
\end{equation}
while band gap due to n=1 resonance corresponds to the frequency range where effective density is negative:
\begin{equation}
\label{eff_n1bandgap}
	\left[\hat{f}^l_{n=1}, \hat{f}^u_{n=1}\right] =
	\left[\frac{\sqrt{2}c_3}{2\pi a} \sqrt{1-\frac{\got{F}\got{K}_{\rho}}{2\left[1-\got{F}(1-\got{K}_{\rho})\right]}}, 
			  \frac{\sqrt{2}c_3}{2\pi a} \sqrt{1+\frac{\got{F\got{K}_{\rho}}}{2\left[1+\got{F}(1-\got{K}_{\rho})\right]}}\right]
\end{equation}
Comparing $n=0$ band gap limits approximated by \eqref{eff_n0bandgap} with those derived using lattice sum expansion \eqref{ps_n0bandgap} it can be seen that they coincide as $\got{F} \rightarrow 0$. On the other hand the limits \eqref{eff_n1bandgap} for the $n=1$ band gap are shifted towards lower frequency compared to the interval \eqref{ps_n1bandgap}. It is noted that as $\got{F} \rightarrow 0$ the approximation of the lower limit $\hat{f}^l_{n=1}$ coincides with the resonance $n=1$ of the elastic shell embedded into the hollow cylinder with rigid walls and radius equal to that of the composite inclusion.

In Figure 4 the dispersion relation \eqref{eff_disprel} is compared with the approximations derived in Sections~\ref{Foldy} and \ref{MAE} in the interval $k_{eff}L \in [0,\pi]$. This interval coincides with that chosen in Section \ref{Foldy} for the wave vector $\vc{\beta}$ (i.e. $\beta L$ belongs to $[0,\pi]$). It is observed that for relatively low filling fractions (see Figure~\ref{fig:CPA}(a) for $\got{F} \approx 0.4$) as well as for high filling fractions (see Figure~\ref{fig:CPA}(b) for $\got{F} \approx 0.6$) the results are accurate to within $5\symbol{37}$ of the semi-analytical solution. It is also noted that as the filling fraction increases the approximation of the upper limit of $n=0$ band gap is less accurate than that obtained with equation \eqref{mae_ulim}.

\begin{figure}[p]
		\center
		\subfigure[]{\includegraphics[scale=1]{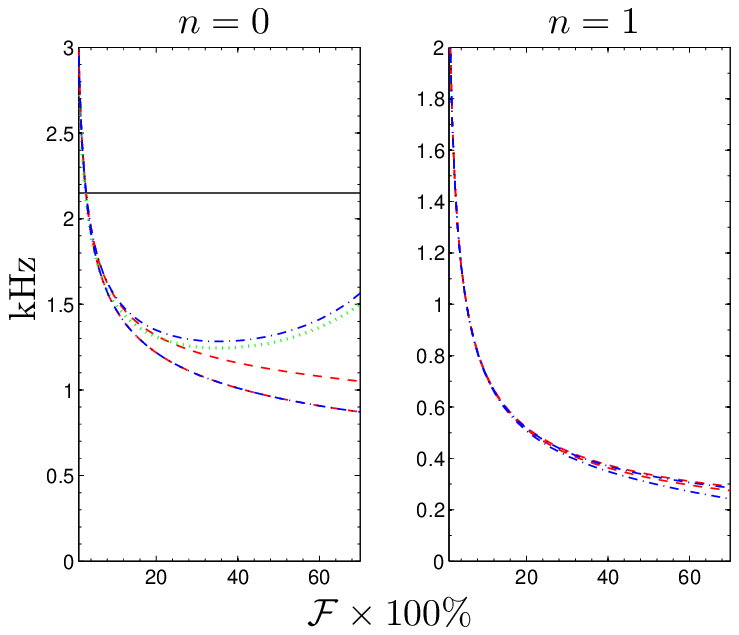}}
		\subfigure[]{\includegraphics[scale=1]{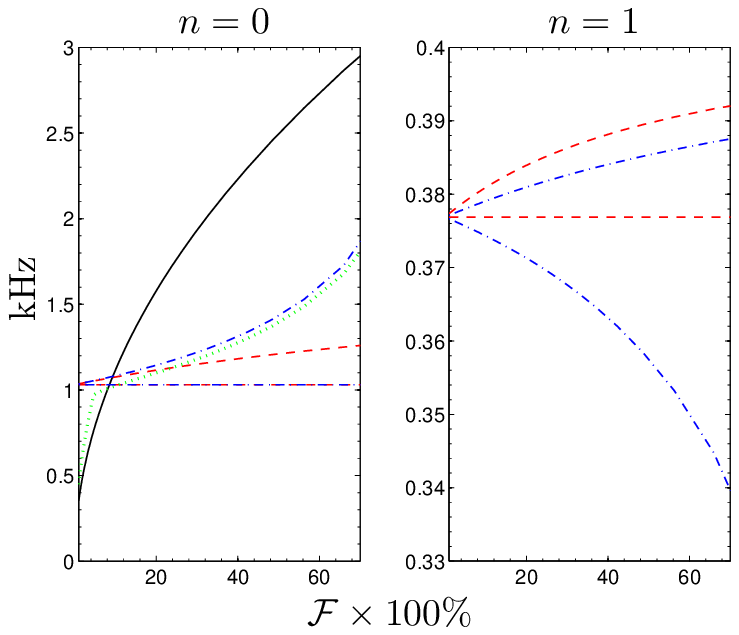}}\\
		\subfigure[]{\includegraphics[scale=1]{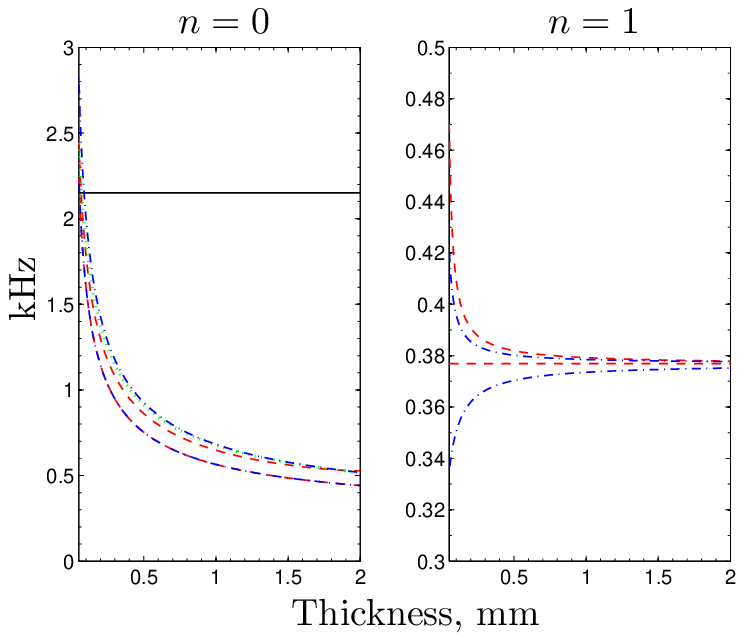}}\hspace{-1mm}
		\subfigure[]{\includegraphics[scale=1]{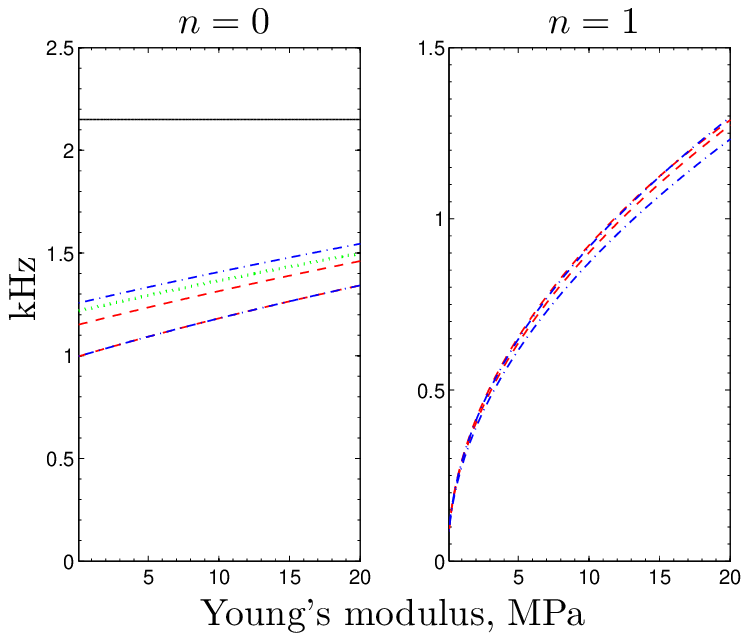}}
    \caption{(Color online) Limits of the band gap $n=0$ and $n=1$ obtained with the Foldy's approximations \eqref{ps_Foldy} ('--~--'), MAE solution \eqref{mae_ulim} ('$\cdot\cdot\cdot$') and self-consistent method \eqref{eff_disprel} ('--~$\cdot$~--'). The first Bragg frequency is plotted with solid line ('--')  (a) Variation of the shell mid-surface radius $a$. (b) Variation of the lattice constant $L$. (c) Variation of the thickness $2h$ with filling fraction $\got{F} \approx 0.4$. (d) Variation of the Young's modulus $E$ with filling fraction $\got{F} \approx 0.4$.}
    \label{fig:Width}
\end{figure}

\section{Results. Band gap width}\label{results}

In Figure~\ref{fig:Width} the results illustrate the application of the approximations derived in the previous sections to estimate the width of the band gaps.

The dependence of the band gaps related to the shell resonances with indices $n=0$ and $n=1$ on the filling fraction is shown in Figures~\ref{fig:Width}(a) and (b). The Bragg frequency $f=c_o/(2L)$ is constant with respect to $\got{F}$ in Figure~\ref{fig:Width}(a) where the characteristic size of the lattice cell (i.e lattice constant $L$) is fixed and the shell radius is varied. It is noted that the approximations based on the Foldy's equation underestimate the upper limit of $n=0$ band gap as $\got{F} \rightarrow 1$. This is improved with the help of the matched asymptotic results and the self-consistent method. It is also observed that for low filling fractions (i.e. $\got{F}<0.1$) the axisymmetric resonance and the corresponding band gap are expected to be observed above the first Bragg frequency where the obtained approximations are not valid. For the band gap attributed to the shell resonance with index $n=1$ the approximations based on the Foldy's equation predicts smaller gap width then that obtained with the self-consistent method. This is due to the fact that the lower limit is fixed on the shell resonance whereas in the self-consistent method it depends on the filling fraction. In Figure~\ref{fig:Width}(b) the shell radius is fixed and the lattice constant is varied. This results in the variation of the Brag frequencies. The obtained results are similar to those in Figure~\ref{fig:Width}(a). Again the derived approximations should accurately predict $n=0$ band gap width. 

It can be observed in Figures~\ref{fig:Width}(a) and (b), that $n=0$ band gap width quickly increases with the filling fraction. For $\got{F}=0.4$ its width is approximately 300 Hz, while for $\got{F}=0.6$ it is 500 Hz. This trend is less pronounced for $n=1$ band gap which remains quite narrow even in dense arrays. For example, the approximations based on the Foldy's equation estimate the band gap width as 10 Hz for $\got{F}=0.4$ and 12 Hz for $\got{F}=0.6$ whereas the self-consistent method gives 20 Hz and 35 Hz, respectively.

In Figure~\ref{fig:Width}(c) the dependence on the shell thickness is illustrated. The variation of the shell thickness also leads to the change in the filling fraction. However, due to the small increment in the outer radius of the elastic shell it is assumed that the filling fraction is fixed (for example, $\got{F}\approx 0.4$ for $2h=2$ mm and $\got{F}\approx 0.37$ for $2h=0.25$ mm). It is expected that the axisymmetric resonance and the corresponding band gap are shifted to the higher frequencies with decrease in the shell thickness. The approximations must become less accurate as the resonance approaches the Bragg frequency so that the results are analysed for $2h>0.1$ mm. The increase in the shell thickness results in the substantial low-frequency shift of $n=0$ band gap. It is predicted that the width of $n=0$ band gap observed around 1.1 kHz is approximately 250 Hz for the shell thickness $2h=0.25$ mm. For the shell thickness $2h=1.5$ mm the band gap is observed around $550$ Hz with width $100$ Hz. It is also noted that the width of $n=1$ band gap increases as the thickness decreases.

The dependence of the upper and lower band gap limits on the elastic modulus of the shell is shown in Figure~\ref{fig:Width}(d). The variation of Young's modulus (i.e. $E\in[0.1,\, 20]$ MPa) results in a shift of the band gap limits. The approximations based on the Foldy's equation underestimate the upper limit of $n=0$ band gap. This is similar to the results shown in Figures~\ref{fig:Width}(a), (b) and (c). It is observed that with decrease of the Young's modulus the band gaps appear at lower frequencies. In the given Young's modulus interval the width of $n=0$ band gap experiences little change (i.e. the width is approximately 250 Hz) whereas $n=1$ band gap width becomes smaller as the Young's modulus decreases. For example, the approximations based on the Foldy's equation estimate the width of $n=1$ band gap as 20 Hz for $E=10$ MPa and $10$ Hz for $E=1$ MPa. The self-consistent method predicts the band gap width double of that obtained with the Foldy's equation.

\section{Conclusions}

In this paper the limits of the band gaps observed in a doubly periodic array of elastic shells and related to the scatterer's resonances are derived. These band gaps are observed below the first Bragg band gap. It is noted that the upper limit of the band gap attributed to the axisymmetric resonance $n=0$ is in the vicinity of the first Bragg band gap that restricts the use of the Foldy's equation. It is shown that this approach can only be used for the substantially small filling fractions (i.e. $<0.4$) when the periodicity has little effect. The improved equation for the upper limit is derived by using the matched asymptotic expansions. This approach includes the periodic effects that results in the accurate approximations for the densely packed periodic array (i.e. $\got{F}>0.4$). In the low-frequency regime the alternative to the Foldy's equation is derived by using the self-consistent method. It is shown that these results can be used for the filling fractions bigger than $0.4$. The methods described here can be easily adapted to derive the analytical expressions for low frequency band gap boundaries in arrays of circular resonant scatterers of different nature (for instance split rings or composite scatterers \cite{SUOU2}).

\section{Acknowledgment}
Part of this work has been supported by EPSRC grants EP/E063136/1 and EP/E062806/1. Authors are grateful for this support. AK is grateful for the support by EU FP7 project, grant agreement no.: 234306.


\end{document}